%% file: Zasche_WASP.tex
\documentclass{elsart}

\usepackage{graphicx,natbib,amssymb,lineno,hyperref}
\usepackage{lscape}

\def\astrobj#1{#1}

\begin{document}

\begin{frontmatter}
\title{Analysis of eight binaries in Lyncis constellation: RV~Lyn, AA~Lyn, AH~Lyn, CD~Lyn, CF~Lyn, DR~Lyn, EK~Lyn, and FS~Lyn.}

\author{P. Zasche}
\ead{zasche@sirrah.troja.mff.cuni.cz}

\address{Astronomical Institute, Faculty of Mathematics and Physics,
 Charles University in Prague, CZ-180 00 Praha 8, V Hole\v{s}ovi\v{c}k\'ach 2, Czech Republic}

\begin{abstract}
The available photometry mainly from the WASP database was used for the first light curve analysis
of eight eclipsing binary systems located in the Lyncis constellation: \astrobj{RV Lyn},
\astrobj{AA Lyn}, \astrobj{AH Lyn}, \astrobj{CD Lyn}, \astrobj{CF Lyn}, \astrobj{DR Lyn},
\astrobj{EK Lyn}, and \astrobj{FS Lyn}. Most of these eclipsing stars are detached ones, having the
orbital periods from 0.54 to 2.3~days. For the systems AA~Lyn and CF~Lyn a non-negligible third
light was detected during the light curve solution. Moreover, 284 new times of minima for these
binaries were derived, trying to identify the period variations. For the system CD~Lyn a
hypothetical third body was detected with the period of about 59~yr.
\end{abstract}

\begin{keyword}
stars: binaries: eclipsing \sep stars: fundamental parameters \PACS 97.10.-q \sep 97.80.-d \sep
97.80.-d
\end{keyword}

\end{frontmatter}

\section{Introduction}

The crucial role of the eclipsing binaries in the nowadays astrophysics is evident. The eclipsing
binary systems (hereafter EB) are being used for the most accurate determination of the basic
parameters such as stellar masses and radii, as distance indicators, or can serve as classical
celestial mechanics laboratories. Nowadays, we can even test the stellar structure models outside
of our Galaxy, see e.g. \cite{2004NewAR..48..731R}. Additionally, also the hidden components can be
studied via the long term observations of the binaries as well as study the dynamical effects in
such multiple systems \citep{2013ApJ...768...33R}. Due to all of these reasons the photometric
monitoring and analysis of the light curves of selected eclipsing binaries still presents a
fruitful contribution to the stellar astrophysics.

However, the available photometry for many interesting eclipsing binaries exists, but some of these
EBs were still not analysed. Therefore, we decided to use mainly the Super WASP photometry
\citep{2006PASP..118.1407P} for a light curve analysis and derivation of new minima times for such
systems, which were not studied before and their light curve solution is missing.

\section{Analysis}

The selection criteria for the binaries included in our study were rather straightforward. We
focused on the unstudied systems in the constellation Lyncis. Only such binaries with known orbital
periods were chosen, having no light curve solution published up to now, have enough data points
for the analysis and also have several published times of minima. The last point was checked via an
online archive of minima times observations, a so-called $O-C$
gateway\footnote{http://var.astro.cz/ocgate/}, see \cite{2006OEJV...23...13P}. Due to the very good
time coverage provided by the Super WASP survey we used this database for the whole analysis of the
light curve. All of the studied systems are located in the Lyncis constellations and are of
moderate brightness (9.3~mag $<$ V $<$ 14.2~mag in maximum) and with the orbital periods ranging
from 0.54 to 2.3 days.

For the analysis of the light curve we used the {\sc PHOEBE} program, ver. 0.31 \citep{Prsa2005},
which is based on the algorithm by \cite{Wilson1971} and its later modifications. Due to having
rather limited information about the stars, some of the parameters have to be fixed for the light
curve (hereafter LC) solution. At first, the "Detached binary" mode (in Wilson \& Devinney mode 2)
was assumed for computing. If some of the components overfills its Roche lobe, we switched to some
other configuration.  The limb-darkening coefficients were interpolated from tables by van~Hamme
(see \citealt{vanHamme1993}), and the linear cosine law was used. The values of the gravity
brightening and bolometric albedo coefficients were set at their suggested values for
convective or radiative atmospheres (see \citealt{Lucy1968}). 
Therefore, the quantities which could be directly calculated from the LC are the
following: the relative luminosities $L_i$, the temperature of the secondary $T_2$,
the inclination $i$, and the Kopal's modified potentials $\Omega_1$ and $\Omega_2$.
The synchronicity parameters $F_1$ and $F_2$ were also fixed at values of 1. The
value of the additional third light contribution $l_3$ was also computed as a free
parameter, which sometimes resulted in a non-negligible value. Its value cannot be
directly compared with the two luminosities $L_1$ and $L_2$, but it is being
compared with the output fluxes $l_1$ and $l_2$ near the quadratures. And finally,
the linear ephemerides were calculated using the available minima times for a
particular system. For the LC modelling, the number of grid points on the
components' surfaces were set to (40,30).

The problem with the mass ratio derivation arose in most of the studied systems. We
started our analysis assuming the mass ratio $q=1$, because no spectroscopy for
these selected systems exists, and for detached EBs the LC solution is almost
insensitive to the photometric mass ratio (see, e.g. \citealt{2005ApSS.296..221T}).
However, because for some systems this approach led to incorrect results, hence we
used an alternative method of deriving the mass ratio following the method e.g. by
\cite{2003MNRAS.342.1334G}. It uses the assumption that both components are located
on the main sequence (which is necessarily not true) and the computed mass ratio is
being directly derived from the individual luminosities. Therefore, having only
limited information in photometry (one filter), some of the individual fitted
parameters suffer from strong correlations between each other. Hence, we have to
emphasize once again that the presented solution is still only very preliminary yet
and only further spectroscopic observations of these systems will be able to reveal
their true nature and the physical parameters with higher conclusiveness.

There also arises the problem with the primary temperature $T_1$. This value has to be fixed during
the whole computing process. When the spectral type is not available, the photometric indices were
used for the rough estimation of the primary temperature (using the tables from
\cite{2013ApJS..208....9P} and the online web
site\footnote{http://www.pas.rochester.edu/$\sim$emamajek/EEM$\_$dwarf$\_$UBVIJHK$\_$colors$\_$Teff.txt}).
However, for some of the systems this approach was not so straightforward due to the fact that for
one star many different photometric indices exist, and moreover their interstellar reddening is not
known. For the systems CD~Lyn, CF~Lyn, DR~Lyn, EK~Lyn, and FS~Lyn some spectral estimations exist
in the literature, while for RV~Lyn, AA~Lyn, and AH~Lyn only the photometric indices are available.
Hence, we collected all available indices and derived the particular spectral estimations. From
these spectral types we eliminated the higher and lower outliers and from the rest some best value
was estimated (or its upper value due to the unknown interstellar extinction).

With the final LC analysis, we also derived many times of minima for a particular system, using a
method as presented in \cite{2014A&A...572A..71Z}. The template of the LC was used to fit the
photometric data from the Super WASP survey. This set of minima times was then combined with the
already published minima mostly taken from the $O-C$ gateway \citep{2006OEJV...23...13P}.

\section{The individual systems}

\subsection{RV Lyn}

The system RV Lyn (also 2MASS J06561142+5051455) is a typical system in our sample of stars. There
were published only a few times of minima for this eclipsing binary and its orbital period of about
2.3~days is known. Nothing more about this system was published, no analysis of its light curve as
well as no spectroscopic study can be found in published papers. We can only roughly estimate its
spectral type from the color indices, hence we fixed the primary temperature at a value of 6700~K
for the whole fitting process.

\begin{figure}
 \includegraphics[width=12cm]{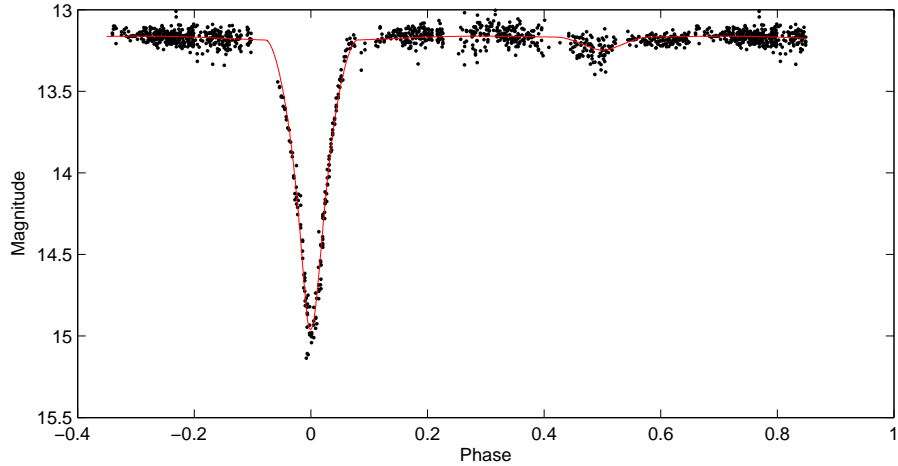}
 \caption{Light curve analysis of RV Lyn, based on the Super WASP photometry.}
 \label{FigRVLynLC}
\end{figure}

The Super WASP photometry revealed that it is a detached system, having very deep primary minimum
(of about 2 magnitudes) and very shallow secondary one. Therefore, the {\sc PHOEBE} code was used
to fit the WASP data and the LC is presented in Fig. \ref{FigRVLynLC}, while the LC parameters are
given in Table \ref{TableLC}. As one can see, both components are rather different from each other
and the primary dominates with its luminosity in the system. No third light was detected in the LC
solution.

\begin{table*}[t]
 \caption{The light-curve parameters as derived from our analysis.}
 \label{TableLC} \centering
 \input{table1.tex}
 \note{$^a$ - given at the orbital phase of 0.25, $^b$ - not fitted during computation.}
\end{table*}

\begin{figure}
 \includegraphics[width=12cm]{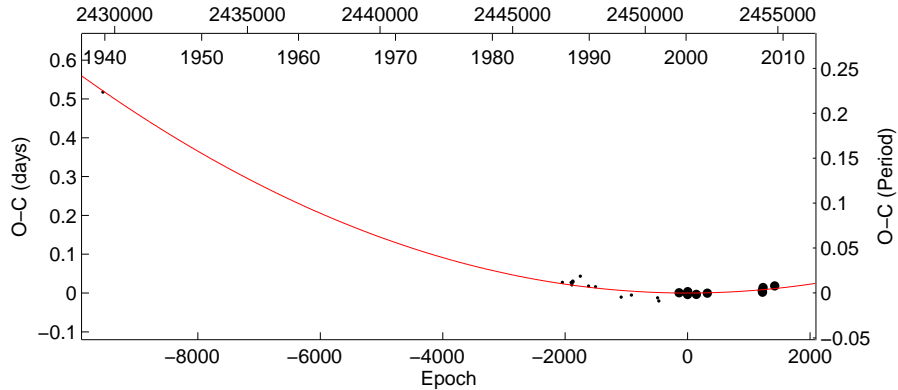}
 \caption{O-C diagram of times of minima for RV Lyn. The black points stand for the
 primary minima, the larger the symbol, the higher the weight. The red line represents the quadratic
 term in the ephemerides.}
 \label{FigRVLynOC}
\end{figure}

Due to very shallow secondary minima, we used only the primary ones for the period analysis.
Despite quite a lot WASP data points only three primary minima have been derived. The resulting
$O-C$ diagram is shown in Fig. \ref{FigRVLynOC}. We can see there that some long-period modulation
of the orbital period is probably present in the system. This can naturally be explained by the
mass transfer between the components, but only further investigation would be able to prove this
hypothesis.

\subsection{AA Lyn}

The eclipsing binary AA Lyn (also 2MASS J07504631+4134065) is rather faint star, which was also not
studied before. \cite{1982AJ.....87..314K} included the star into their survey of RR~Lyrae stars,
but with the note that it is an eclipsing binary with the period of about 0.56~days. Since then no
analysis of AA~Lyn was carried out.

\begin{figure}
 \includegraphics[width=12cm]{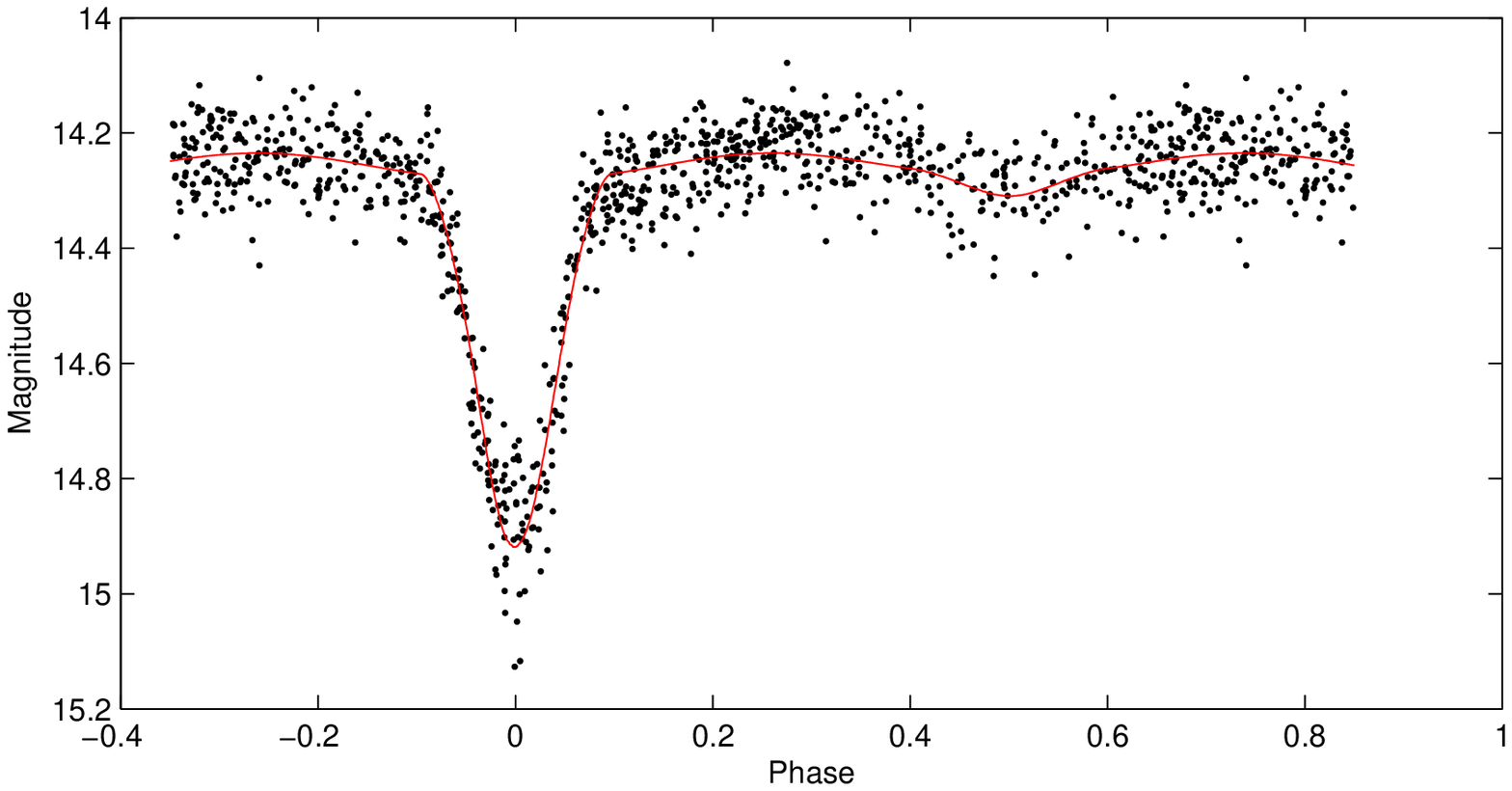}
 \caption{Light curve analysis of AA Lyn, based on the Super WASP photometry.}
 \label{FigAALynLC}
\end{figure}

For the LC fitting we assumed the primary temperature to be of 5600~K (it is the coolest star in
our sample) and used the WASP photometry for the LC analysis. The result is plotted in Fig.
\ref{FigAALynLC}, and the LC parameters as resulted from {\sc PHOEBE} are given in Table
\ref{TableLC}. This is the only system which resulted in semidetached configuration. One can see
that the primary is the dominant source in the system, but there also arose a non-negligible
contribution of the third light. However, its origin is still questionable because there are two
close companions to AA~Lyn (see \citealt{2000A&AS..143...33B}) at the distances of a few arcseconds
only.

\begin{figure}
 \includegraphics[width=12cm]{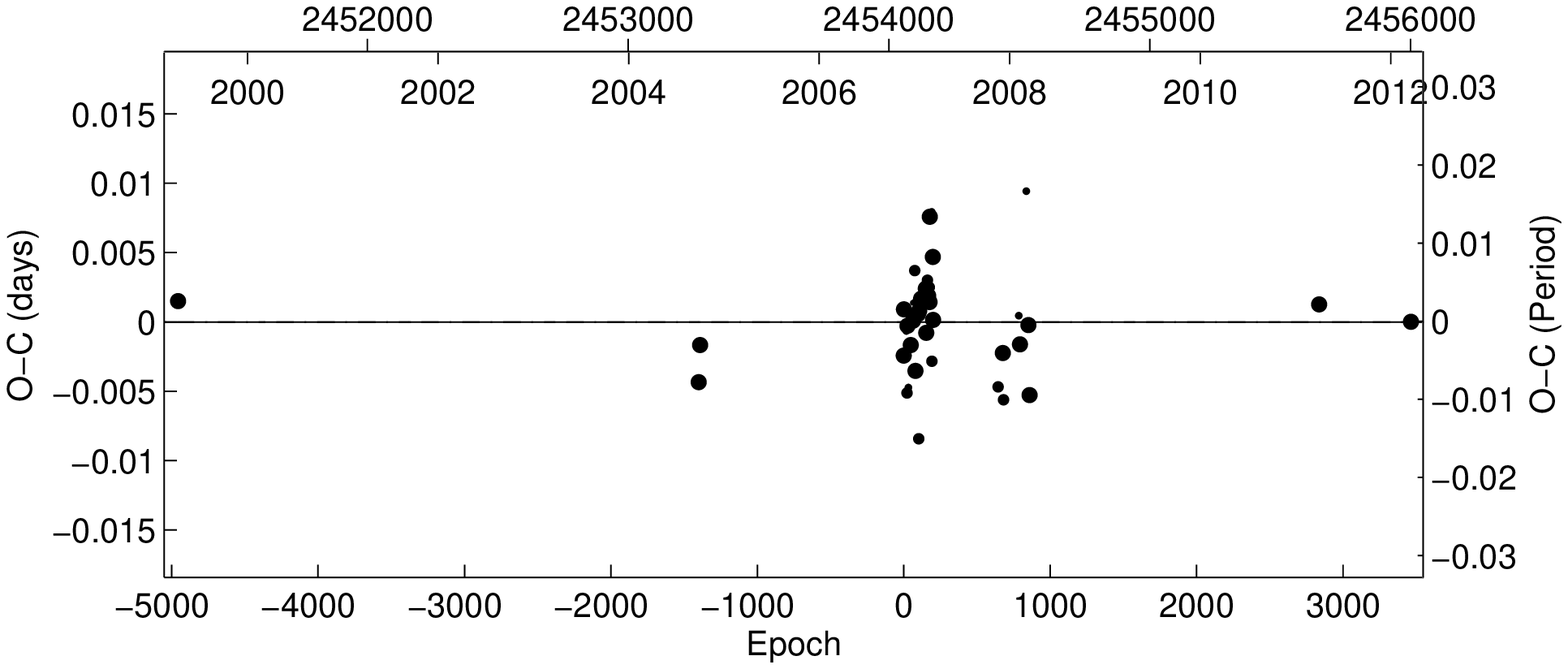}
 \caption{O-C diagram of times of minima for AA Lyn.}
 \label{FigAALynOC}
\end{figure}

Because of shallow secondary minimum, only the primary ones were used for a period analysis. In our
Fig. \ref{FigAALynOC} there are plotted the new times of minima together with the already published
ones. Obviously, there is no variation in the times of minima, or our dataset is still too poor for
any such detection.

\subsection{AH Lyn}

The system AH Lyn (also 2MASS J08421824+3711051) is the binary which was also not studied before,
therefore we included it into our sample of stars. AH~Lyn was included into the study of RR~Lyrae
stars \citep{1982AJ.....87..314K} like AA~Lyn and the authors correctly derived its orbital period
to be of about 1.016~days. Since then only several publications with the times of minima were
published.

\begin{figure}
 \includegraphics[width=12cm]{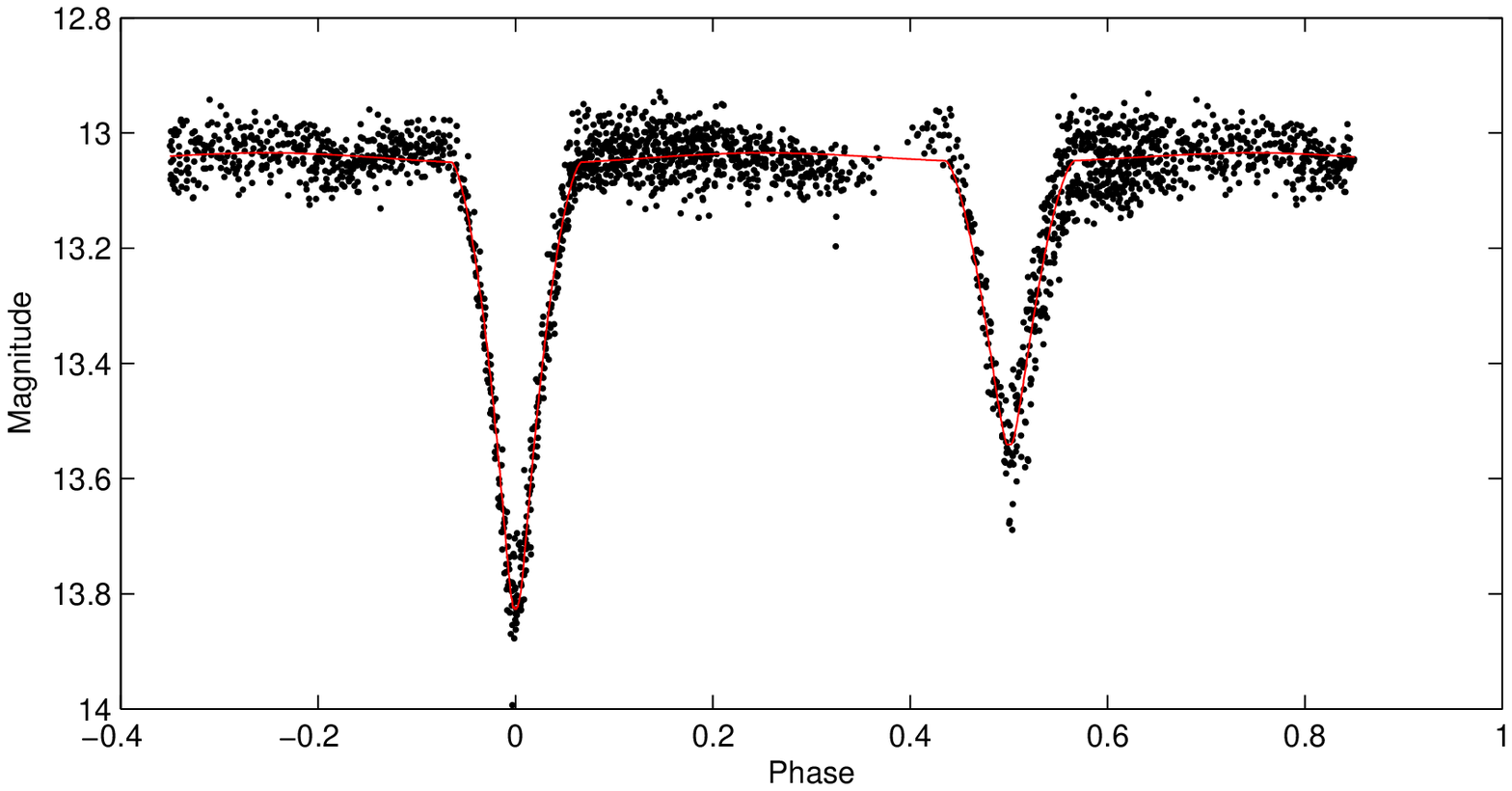}
 \caption{Light curve analysis of AH Lyn, based on the Super WASP photometry.}
 \label{FigAHLynLC}
\end{figure}

For the light curve analysis we fixed the primary temperature to $T_1=6500$~K in agreement with the
photometric indices of the star. The WASP photometry shows us that both the eclipses are rather
deep and symmetrically shaped. The final parameters of the LC fitting are given in Table
\ref{TableLC}, while the LC plot is shown in Fig. \ref{FigAHLynLC}. The system is well detached,
both components are rather similar to each other, and no third light was detected in the LC
solution.

\begin{figure}
 \includegraphics[width=12cm]{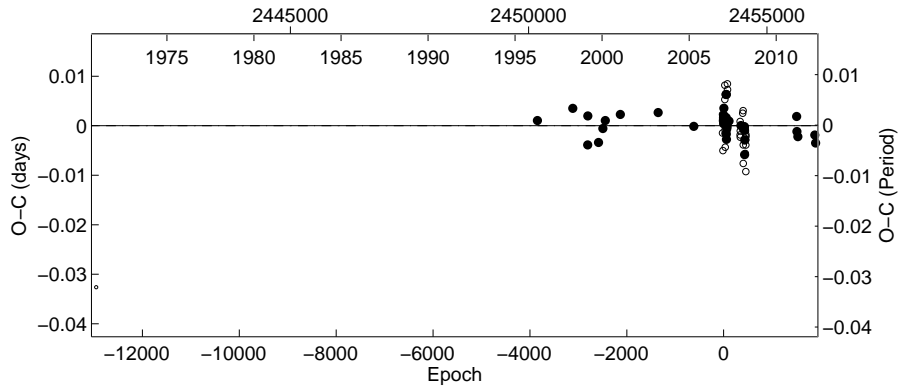}
 \caption{O-C diagram of times of minima for AH Lyn. The secondary minima are plotted
 as open circles.}
 \label{FigAHLynOC}
\end{figure}

For the period analysis we derived 60 new minima times (both primary and secondary) from the WASP
data covering almost 500 days. With the already published ones the complete dataset is plotted in
Fig. \ref{FigAHLynOC}, but no visible variation can be seen there.

\subsection{CD Lyn}

The star named CD Lyn (also HIP 37615) is relatively bright star, and it is also the most
frequently studied one. There exist two dedicated studies on CD Lyncis, \cite{2000IBVS.4911....1B}
and \cite{2002BAVSR..51....5M}, but these are only remarks on their observations of CD Lyn
photometrically, with no LC analysis. Moreover, \cite{2000IBVS.4911....1B} presented the orbital
period of 4.549~days, abandoning the original 2.27~days period, arguing that there is no curvature
near the quadrature. But as we can see from our analysis, the correct period is 2.27~days for sure.

\begin{figure}
 \includegraphics[width=12cm]{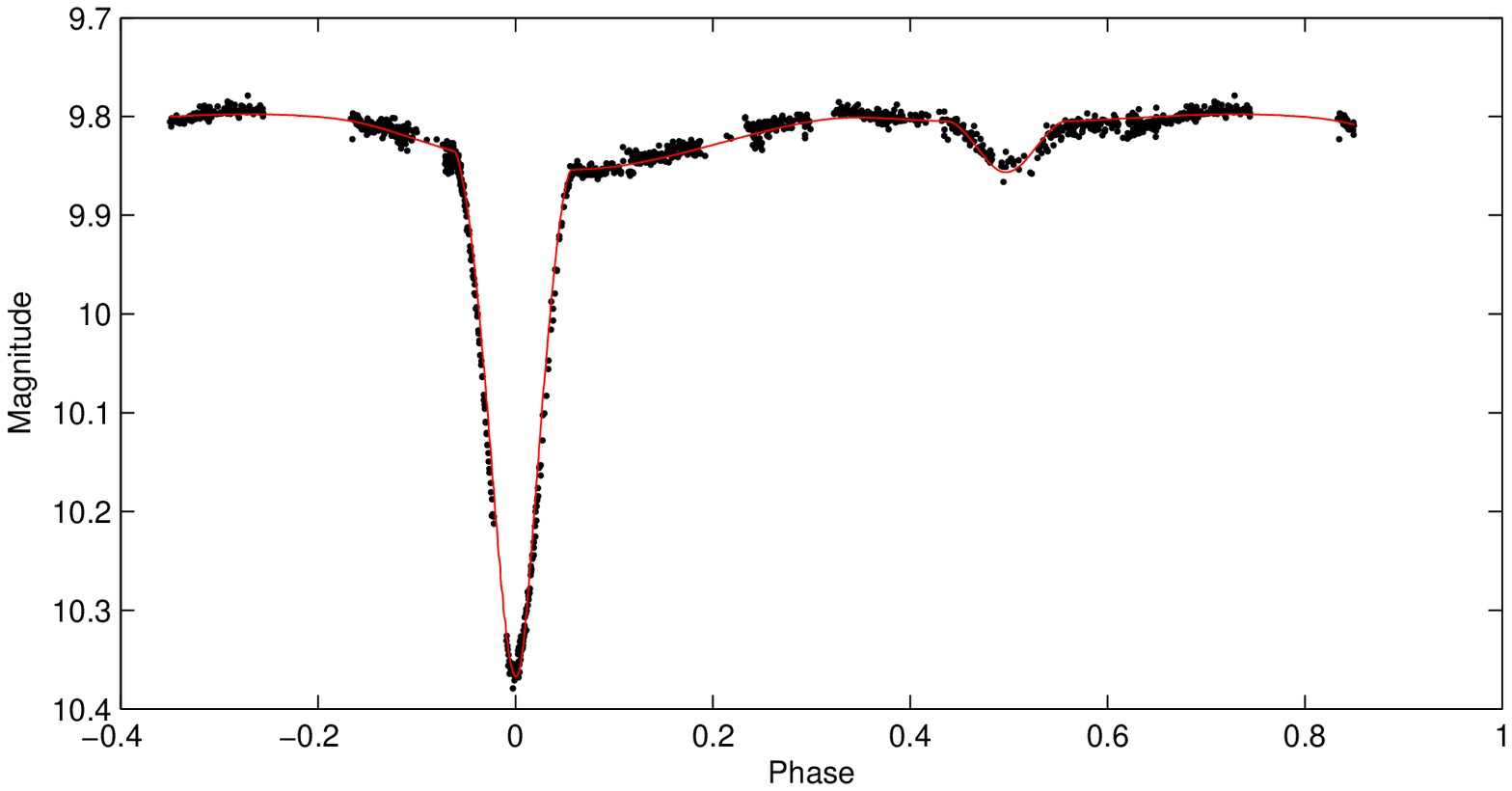}
 \caption{Light curve analysis of CD Lyn, based on the Super WASP photometry.}
 \label{FigCDLynLC}
\end{figure}

\begin{figure}
 \includegraphics[width=12cm]{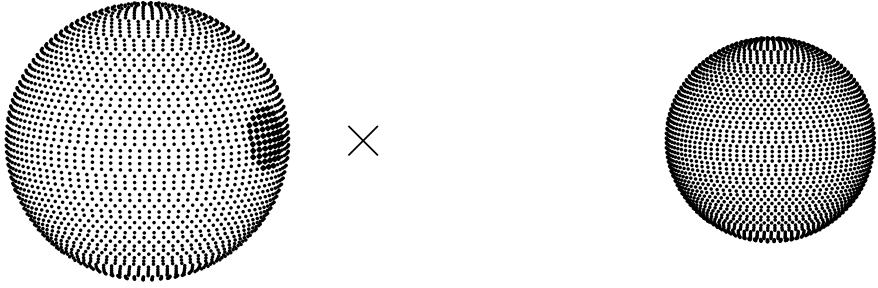}
 \caption{3D plot of CD Lyn, cross stands for the barycenter of the system.}
 \label{FigCDLyn3D}
\end{figure}

The LC fitting of the Super WASP data using the {\sc PHOEBE} programme was using the assumption of
$T_1 = 6800~$K, because its spectral type was derived as F2 by \cite{1952CoRut..32....1H}. The
final fit of the LC is given in Fig. \ref{FigCDLynLC}, and the LC parameters are written in Table
\ref{TableLC}. As one can see, the secondary minima have much less depth, but definitely cannot be
taken as a noise. Hence, the main finding as published by \cite{2000IBVS.4911....1B} has to be
reconsidered. Another interesting finding about this star is the asymmetric shape of its light
curve, hence we have to use a hypothesis of a star spot on the surface of primary (see Fig.
\ref{FigCDLyn3D}). However, it seems like the shape of the LC is changing in time, maybe due to the
moving spot or some other photospheric activity of the star(s).

\begin{figure}
 \includegraphics[width=12cm]{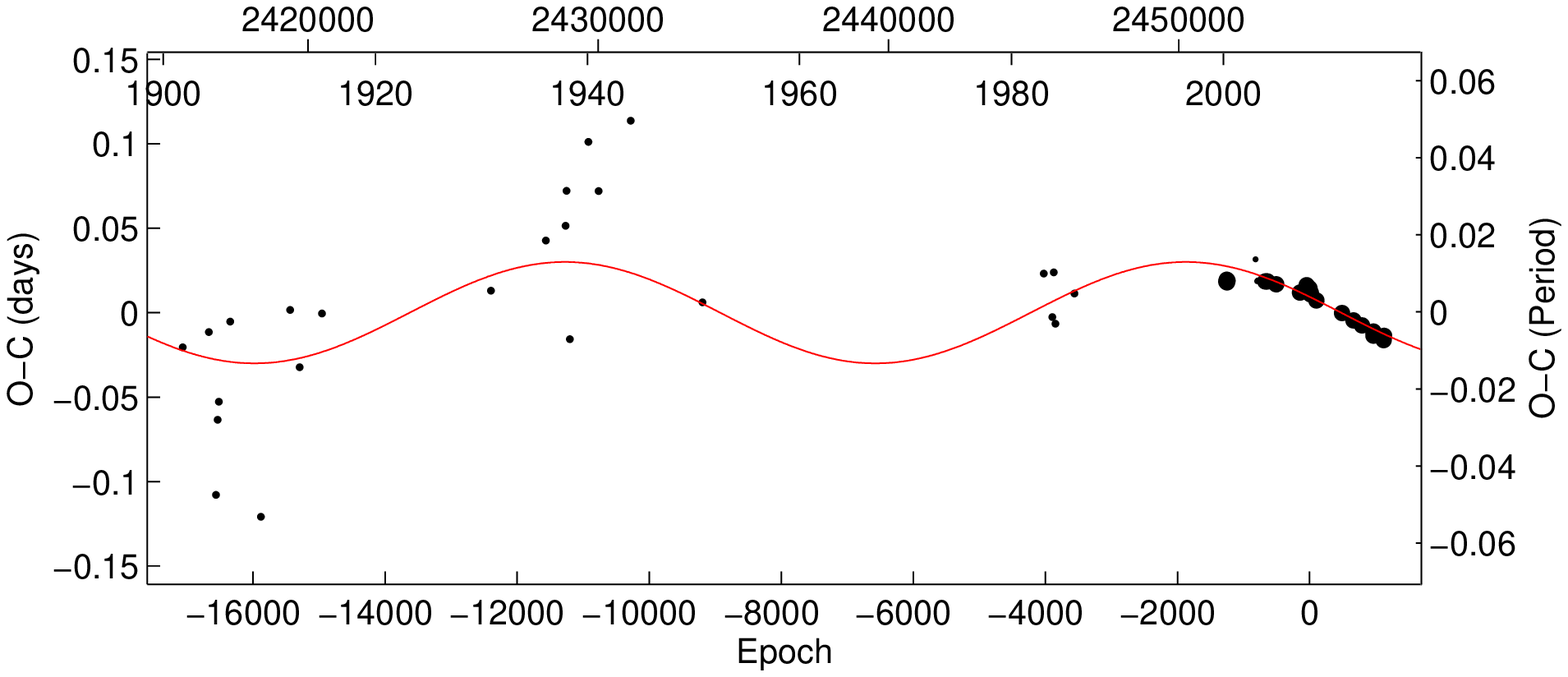}
 \caption{O-C diagram of times of minima for CD Lyn.}
 \label{FigCDLynOC}
\end{figure}

The period analysis of CD~Lyn was done using the already published data as well as our new data
points (altogether eight new primary minima from the WASP survey). The result is shown in Fig.
\ref{FigCDLynOC}, where we also used the hypothesis of the third body orbiting around a common
barycenter with the eclipsing pair (see e.g. \citealt{Irwin1959} or \citealt{Mayer1990}). This
approach was used because it produces much better result than the linear or quadratic ephemerides
term for description of the period variation. The variation of such a component has the period of
about 59~years and the amplitude of about 0.03~days in the $O-C$ diagram. From our LITE fit we also
predicted that such a body should present at least of about 2.5\% of the total luminosity, but our
LC solution results in zero value. Therefore, for a final confirmation of any such body in the
system one needs much more data, so our presented solution is still just a hypothesis yet.

\subsection{CF Lyn}

Another rather bright target is CF Lyn (also HIP 37748), which has the orbital period of about
1.4~days, but was also not studied in detail neither photometrically, nor spectroscopically. Only
its spectral type was classified as F8 by \cite{1975ascp.book.....H}.

\begin{figure}
 \includegraphics[width=12cm]{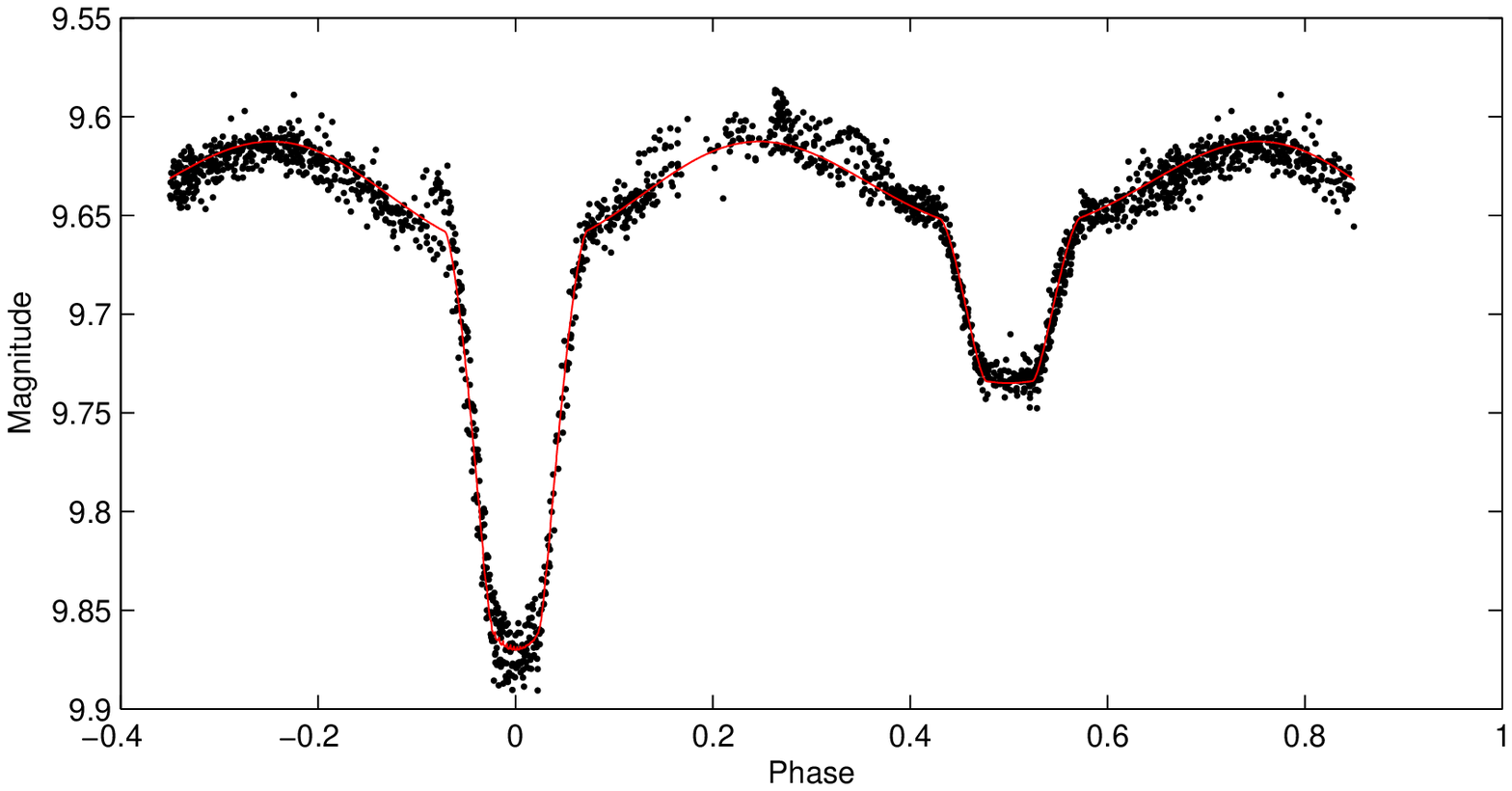}
 \caption{Light curve analysis of CF Lyn, based on the Super WASP photometry.}
 \label{FigCFLynLC}
\end{figure}

We used the WASP photometry for the light curve modelling and the assumption of the
6150~K for the primary temperature. As we can see from Fig. \ref{FigCFLynLC}, the
star has relatively shallow total eclipses, which could indicate large fraction of
the third light and inclination close to 90$^\circ$. For this system we also tried a
different approach for the analysis. Due to its significant curvature outside of
eclipses we also tried to fit the mass ratio $q$ as a free parameter despite its
detached configuration. This result was then compared with the result as obtained
via a standard method of $q$ estimation by a Graczyk's method. And both $q$
parameters resulted in rather similar values of about 0.85 and 0.83, respectively.
All the parameters of our LC fitting are given in Table \ref{TableLC2} (a solution
with fitted $q$ is presented). Rather significant value of the third light resulted,
indicating possible presence of the third component in the system. Noticeable is
also some light curve variability over the Super WASP data period.

\begin{figure}
 \includegraphics[width=12cm]{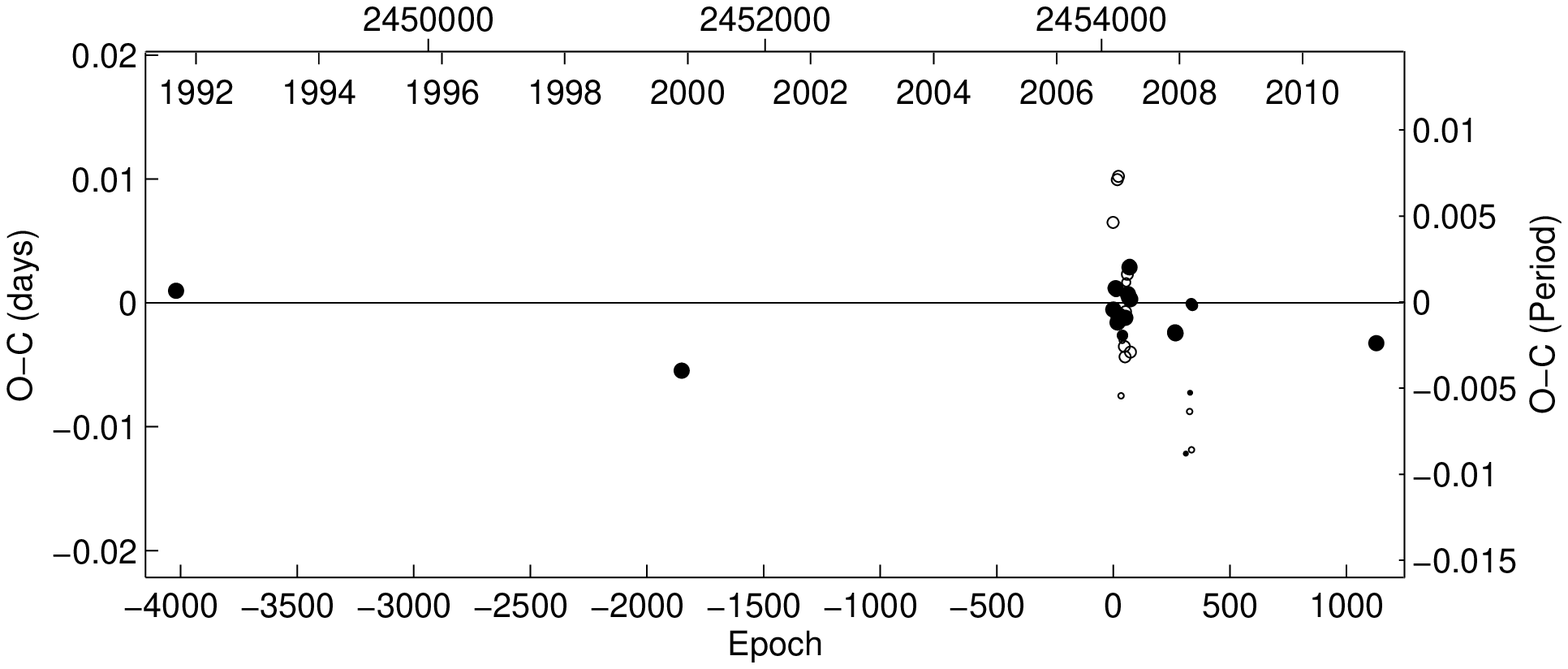}
 \caption{O-C diagram of times of minima for CF Lyn.}
 \label{FigCFLynOC}
\end{figure}

Concerning the period analysis we collected only the three minima as presented in the $O-C$ gateway
\citep{2006OEJV...23...13P} and together with our 33 new times of minima, we have the coverage over
almost 20 years of data. However, even on this dataset there is no evident variation of the period,
see Fig. \ref{FigCFLynOC}.

\begin{table*}[t]
 \caption{The light-curve parameters as derived from our analysis.}
 \label{TableLC2} \centering
 \input{table2.tex}
 \note{$^a$ - given at the orbital phase of 0.25}
\end{table*}

\subsection{DR Lyn}

The star DR Lyn (also TYC 3421-2216-1) is another Algol-type eclipsing binary in our sample of
stars. No detailed study about this star was found in the published papers, only three times of
minima were published till yet. These minima gave the orbital period of about 1.78~days. The star
was also included into the survey of LAMOST \citep{2015RAA....15.1095L}, yielding its spectral type
of about F3 and the primary temperature to be 6690~K.

\begin{figure}
 \includegraphics[width=12cm]{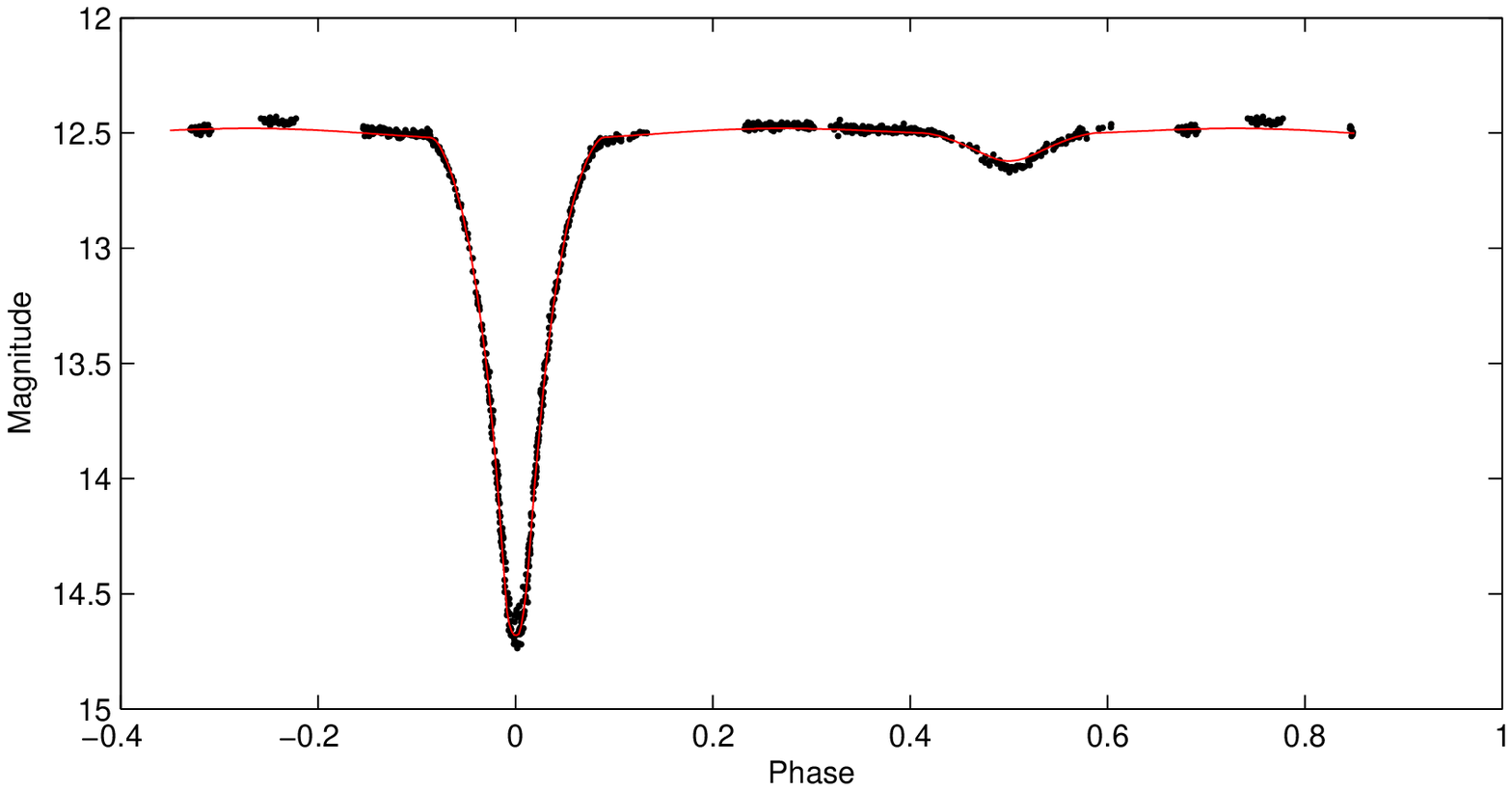}
 \caption{Light curve analysis of DR Lyn, based on the Super WASP photometry.}
 \label{FigDRLynLC}
\end{figure}

With this assumed temperature we performed the light curve analysis of the WASP data. The light
curve shape plotted in Fig. \ref{FigDRLynLC} shows very deep primary minimum (more than 2
magnitudes) and rather shallow secondary one. It indicates quite different components in the
eclipsing pair. The results of the LC modelling are given in Table \ref{TableLC2}. The components
are well detached, but rather different from each other.

\begin{figure}
 \includegraphics[width=12cm]{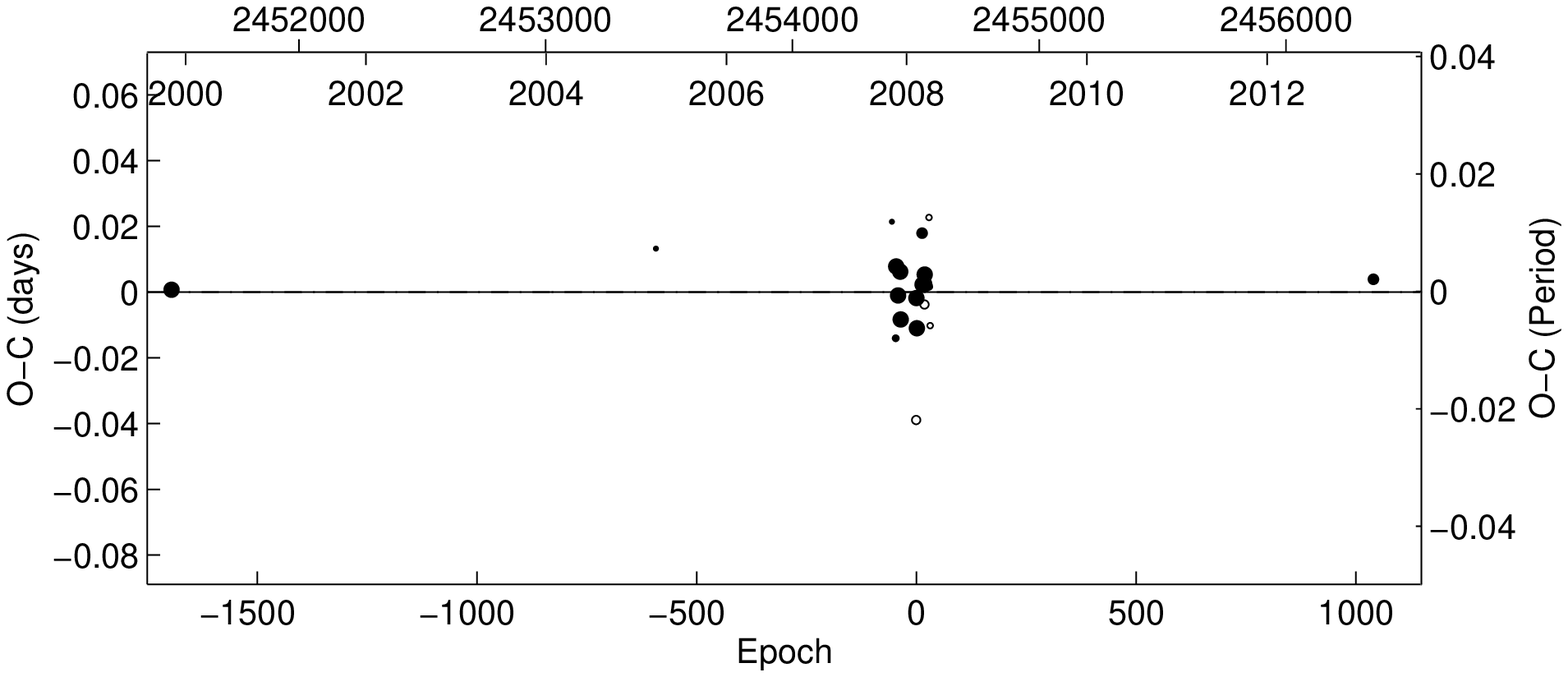}
 \caption{O-C diagram of times of minima for DR Lyn.}
 \label{FigDRLynOC}
\end{figure}

For the period analysis we collected the already published data points (i.e. only three times of
minima) together with our new ones from the WASP survey (i.e. 19 new minima). With this dataset we
carried out the analysis, but no periodic signal was found, see our Fig. \ref{FigDRLynOC}.

\subsection{EK Lyn}

The star EK Lyn (also TYC 2973-339-1) is the brightest target in our sample, having the orbital
period of about 2.23~days. Despite its high luminosity, only very little is known about this star.
The only relevant information is that one by \cite{1975ascp.book.....H} that the spectral type is
of A2, hence it is the system of the earliest spectral type in our sample of stars.

\begin{figure}
 \includegraphics[width=12cm]{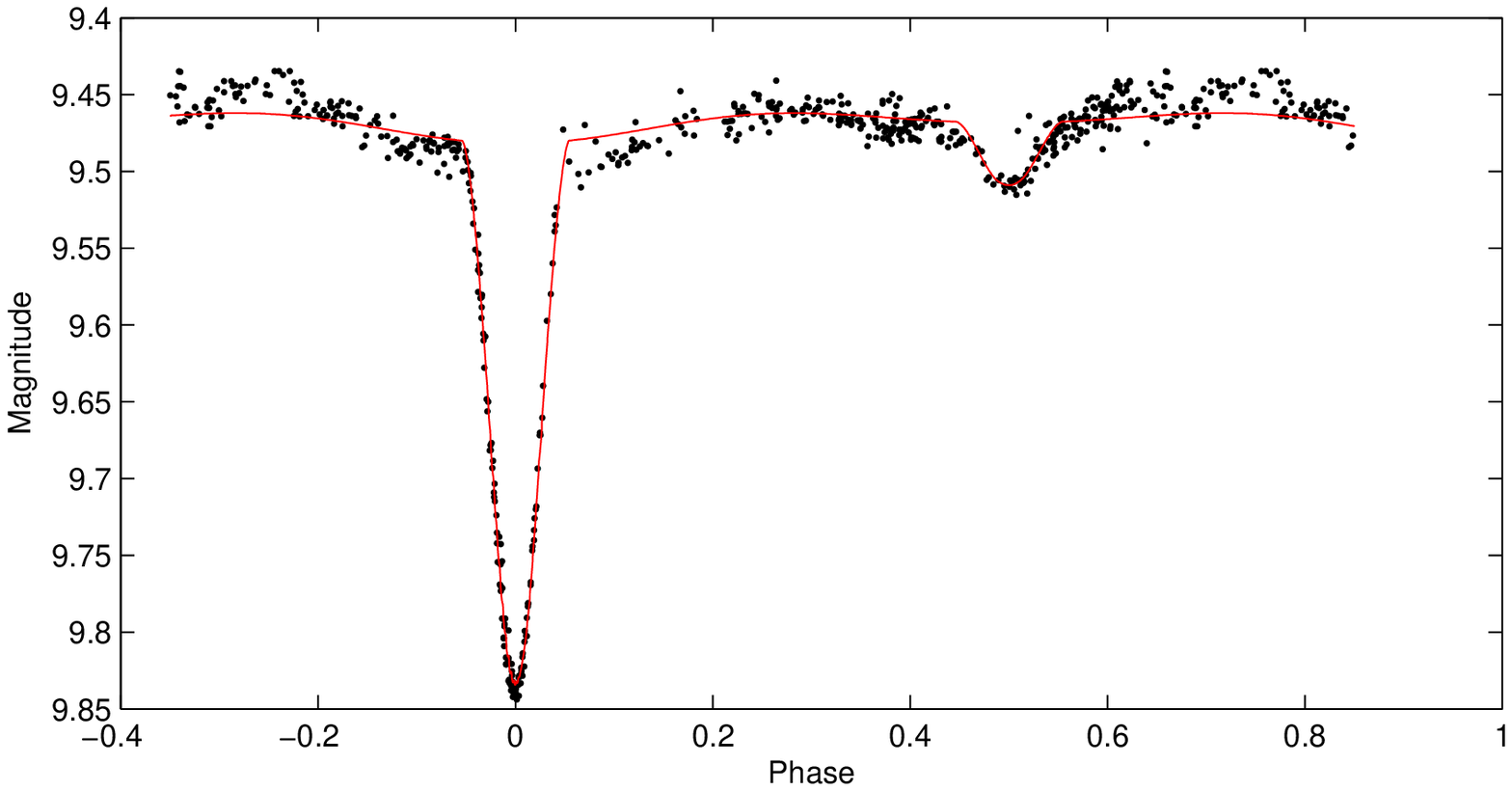}
 \caption{Light curve analysis of EK Lyn, based on the Super WASP photometry.}
 \label{FigEKLynLC}
\end{figure}

For the light curve analysis we fixed the primary temperature to 8840~K (in agreement with
\citealt{2013ApJS..208....9P}). The shape of the LC is changing in time, hence its modelling was
not straightforward. The parameters of our LC fit are given in Table \ref{TableLC2}, where we can
see that the primary component dominates the system and no third light was detected. The fit is
also plotted in Fig. \ref{FigEKLynLC}.

\begin{figure}
 \includegraphics[width=12cm]{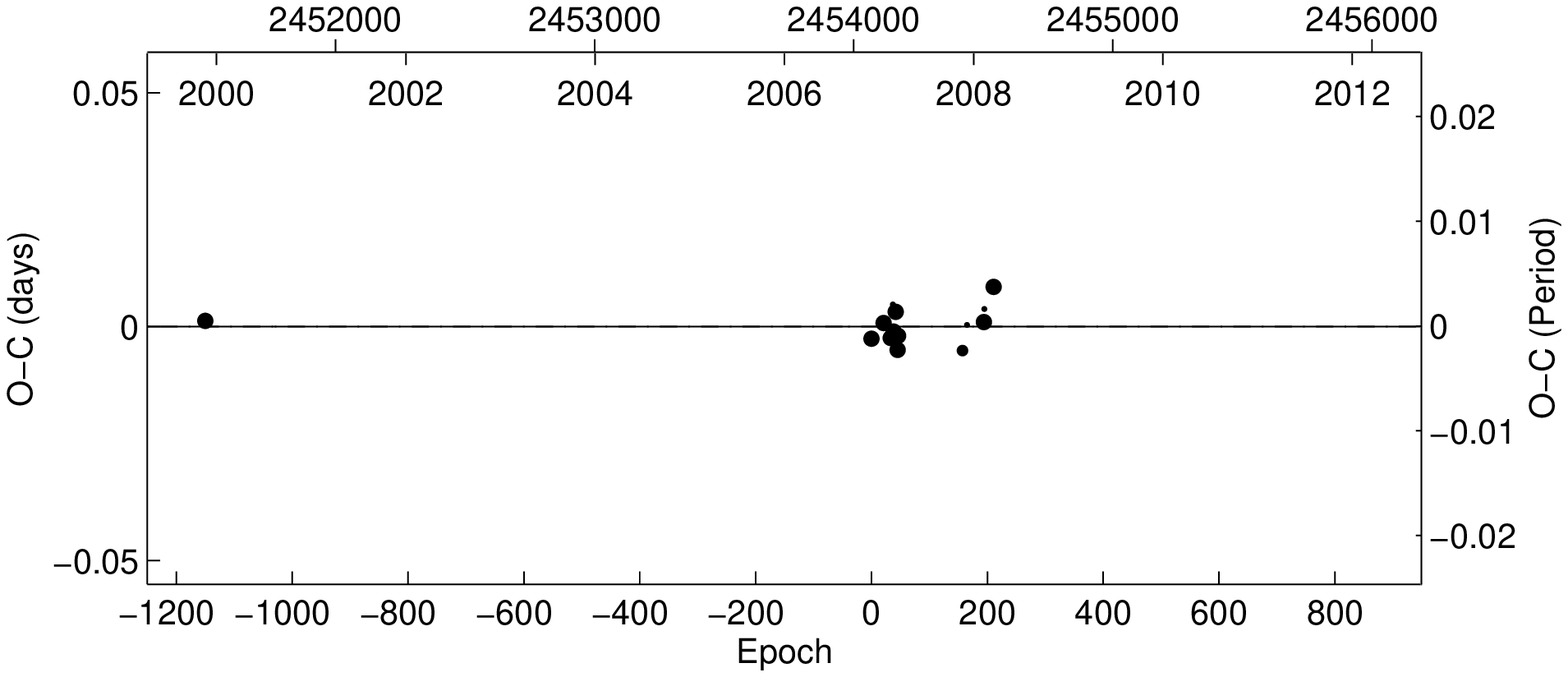}
 \caption{O-C diagram of times of minima for EK Lyn.}
 \label{FigEKLynOC}
\end{figure}

Analysis of the period changes was done using the two published minima together with our new 13
data points. The result is shown in Fig. \ref{FigEKLynOC}. No visible variation can be seen there,
but the data set is still rather limited yet. The most recent minima deviate a bit from the linear
ephemerides, but only further investigation would prove any such variation. What is quite
surprising is the fact that the time of minimum published by \cite{2012IBVS.6029....1D} deviates of
about 0.5 days from our ephemerides and is probably incorrect (we have not plotted this one data
point in our Fig. \ref{FigEKLynOC}).

\subsection{FS Lyn}

The eclipsing system FS Lyn (also TYC 2986-534-1) is the only one system in our study, which was
classified as a $\beta$-Lyrae type star \citep{2005BaltA..14..205M}. However, despite its short
period (0.54~days) and relatively high brightness (11~mag) it was not studied before.

\begin{figure}
 \includegraphics[width=12cm]{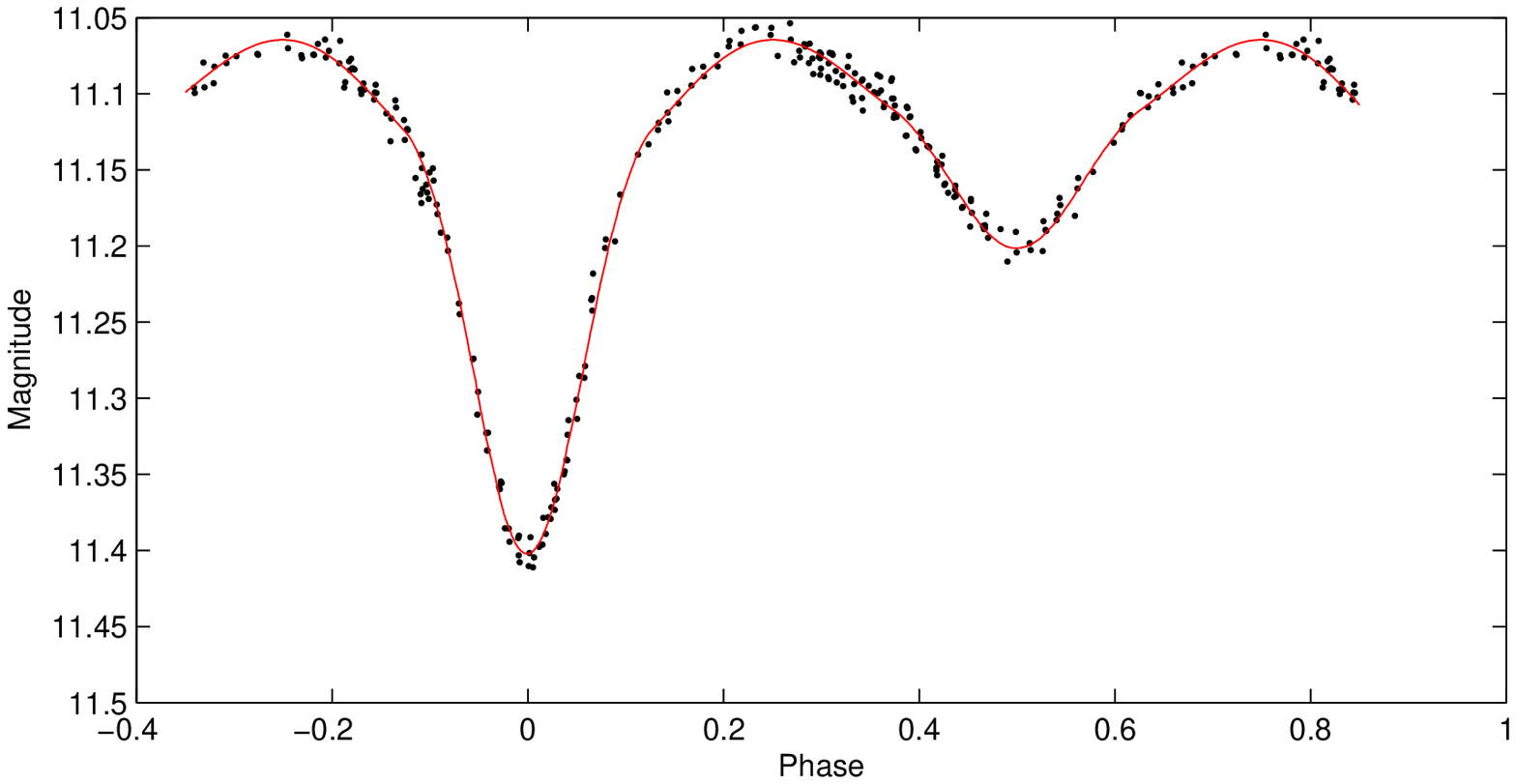}
 \caption{Light curve analysis of FS Lyn, based on the Super WASP photometry.}
 \label{FigFSLynLC}
\end{figure}

For the LC modelling we used the assumption that the star is of about F2 spectral
type, i.e. the primary temperature was fixed at a value of 7100~K, see
\cite{2006ApJ...638.1004A}. Performing the LC modelling, we obtained a solution with
a detached configuration (near contact, but both semidetached and contact
configurations were tested but produced slightly worse fits). Due to its shape of
the light curve, we also tried to compute the mass ratio as a free parameter. There
resulted that the $q$ values from the classical approach of Graczyk and that one
fitted are not so different from each other (0.71 fitted, while 0.64 estimated from
the mass-luminosity relation method by \citealt{2003MNRAS.342.1334G}). The resulting
parameters are given in Table \ref{TableLC2}, while the fit itself is plotted in
Fig. \ref{FigFSLynLC}.

\begin{figure}
 \includegraphics[width=12cm]{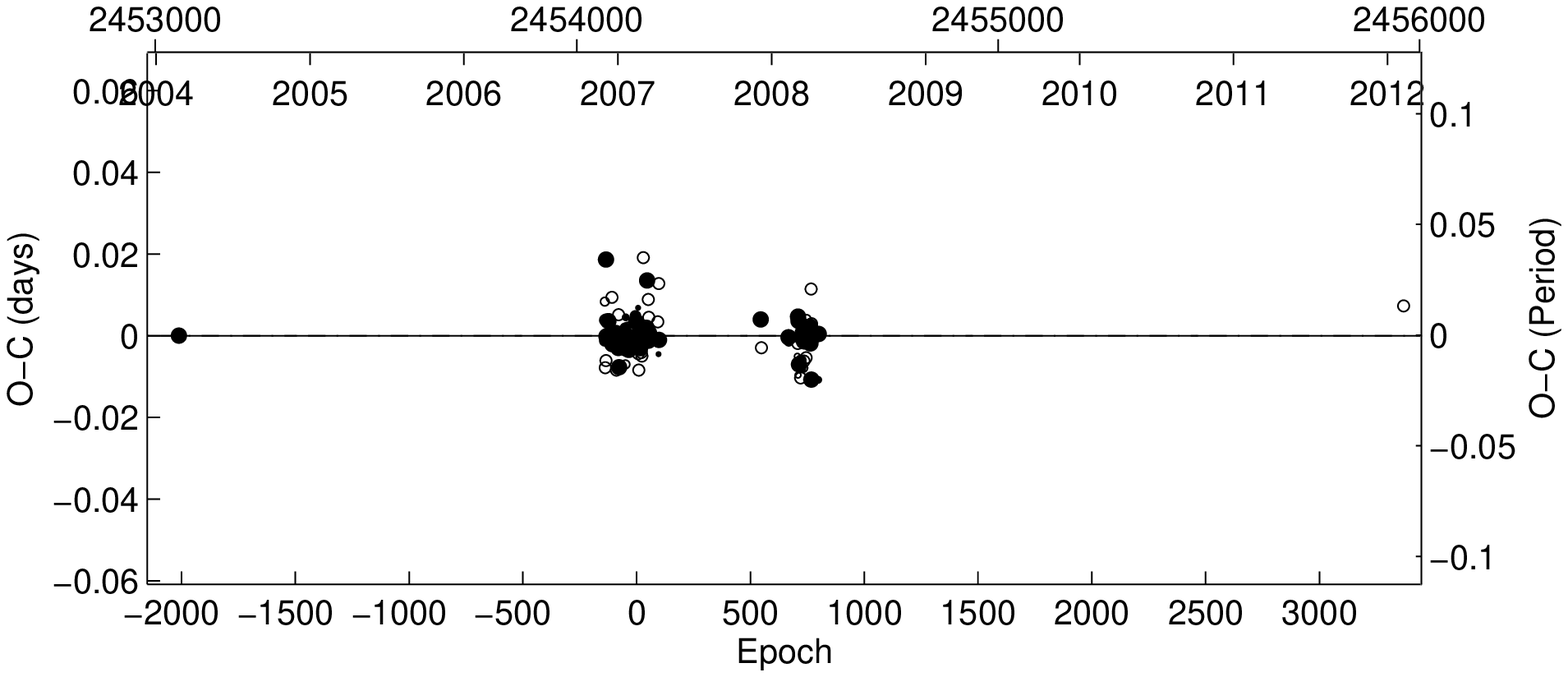}
 \caption{O-C diagram of times of minima for FS Lyn.}
 \label{FigFSLynOC}
\end{figure}

For the analysis of its orbital period we derived from the WASP data 107 minima in total. With the
two already published ones the complete data set covers about 8 years, see Fig. \ref{FigFSLynOC}.
However, no variation is visible on these data and the linear ephemerides are sufficient for
prospective future observations.

\section{Discussion and conclusions}

The very first LC solution for eight Algol-type eclipsing binaries (based on the Super WASP
photometry) led to several findings:

\begin{itemize}
  \item The photometry based on the Super WASP survey data can be used for a fruitful analysis for the
  eclipsing binaries never studied before.
  \item Second-order effects such as the third light or the spots, are also detectable in these data.
  \item For two of the systems (AA Lyn, and CF~Lyn) the amount of the third light is large enough that
  these cannot easily be considered as pure binaries in any future more detailed study.
  \item The method of using the light curve templates for deriving the times of minima provides us
  with reliable and sufficiently precise times of minima suitable for a period analysis.
  \item For the system RV~Lyn we found a steady period increase (probably due to mass transfer), while
  for the system CD~Lyn there was detected some period modulation in the $O-C$ diagram. This variation
  with the period of about 59 years can be attributed to a prospective third body in the system.
\end{itemize}

All of the presented systems have not been studied before concerning their light curves, hence we
can consider this study as a good starting point for a future more detailed analysis. Particularly,
a special focus should be take to these systems, where a larger fraction of the third light was
detected and the system, where a third body variation in the $O-C$ diagram was found.

\section{Acknowledgments}
We thank the "Super WASP" team for making all of the observations easily public available. This
paper makes use of data from the DR1 of the WASP data \citep{2010A&A...520L..10B} as provided by
the WASP consortium, and the computing and storage facilities at the CERIT Scientific Cloud, reg.
no. CZ.1.05/3.2.00/08.0144 which is operated by Masaryk University, Czech Republic. This
investigation was supported by the Czech Science Foundation grant no. GA15-02112S. This research
has made use of the SIMBAD database, operated at CDS, Strasbourg, France, and of NASA's
Astrophysics Data System Bibliographic Services.

\newpage

\begin{table}
 \caption{New Super WASP heliocentric minima times for the studied systems.}
 \label{TableMin1} \centering \scalebox{0.68}{
 \tiny
\begin{tabular}{ c c c c | c c c c | c c c c | c c c c}
\hline \hline
     Star  &  HJD        & Error   & Type  &      Star  &  HJD        & Error   & Type &  Star  &  HJD        & Error   & Type  &      Star  &  HJD        & Error   & Type\\
           & 2400000+    & [days]  &       &            & 2400000+    & [days]  &      &        & 2400000+    & [days]  &       &            & 2400000+    & [days]  &     \\ \hline
RV Lyn & 54409.55525 & 0.00107 & Pri & AH Lyn & 54148.46471 & 0.00062 & Pri & CF Lyn & 54532.41182 & 0.00035 & Pri & FS Lyn & 54140.54703 & 0.00024 & Sec  \\
RV Lyn & 54418.78147 & 0.00105 & Pri & AH Lyn & 54149.47737 & 0.00066 & Pri & CF Lyn & 54534.47809 & 0.00025 & Sec & FS Lyn & 54141.35882 & 0.00040 & Pri  \\
RV Lyn & 54439.56228 & 0.01397 & Pri & AH Lyn & 54150.49752 & 0.00053 & Pri & CF Lyn & 54539.33855 & 0.00140 & Pri & FS Lyn & 54141.62519 & 0.00046 & Sec  \\
AA Lyn & 53270.71845 & 0.00081 & Pri & AH Lyn & 54153.54617 & 0.00035 & Pri & DR Lyn & 54402.86409 & 0.00187 & Pri & FS Lyn & 54142.43395 & 0.00005 & Pri  \\
AA Lyn & 53275.77618 & 0.00080 & Pri & AH Lyn & 54154.56193 & 0.00057 & Pri & DR Lyn & 54418.85657 & 0.00280 & Pri & FS Lyn & 54143.51490 & 0.00036 & Pri  \\
AA Lyn & 54056.66579 & 0.00044 & Pri & AH Lyn & 54155.57885 & 0.00035 & Pri & DR Lyn & 54420.65931 & 0.00107 & Pri & FS Lyn & 54145.40421 & 0.00039 & Sec  \\
AA Lyn & 54057.78959 & 0.00033 & Pri & AH Lyn & 54156.59493 & 0.00035 & Pri & DR Lyn & 54427.77398 & 0.00013 & Pri & FS Lyn & 54145.68229 & 0.00032 & Pri  \\
AA Lyn & 54066.77067 & 0.00043 & Pri & AH Lyn & 54157.61106 & 0.00055 & Pri & DR Lyn & 54436.68563 & 0.00051 & Pri & FS Lyn & 54146.48110 & 0.00026 & Sec  \\
AA Lyn & 54069.57356 & 0.00037 & Pri & AH Lyn & 54169.29889 & 0.00051 & Sec & DR Lyn & 54437.57588 & 0.00023 & Sec & FS Lyn & 54146.75387 & 0.00068 & Pri  \\
AA Lyn & 54070.70159 & 0.00047 & Pri & AH Lyn & 54170.32503 & 0.00014 & Sec & DR Lyn & 54438.45196 & 0.00212 & Pri & FS Lyn & 54147.55707 & 0.00074 & Sec  \\
AA Lyn & 54074.62396 & 0.00048 & Pri & AH Lyn & 54171.34025 & 0.00043 & Sec & DR Lyn & 54501.64267 & 0.00101 & Sec & FS Lyn & 54149.45652 & 0.00017 & Pri  \\
AA Lyn & 54083.61287 & 0.00061 & Pri & AH Lyn & 54172.34865 & 0.00103 & Sec & DR Lyn & 54502.57016 & 0.00046 & Pri & FS Lyn & 54150.53247 & 0.00021 & Pri  \\
AA Lyn & 54084.73688 & 0.00021 & Pri & AH Lyn & 54436.61451 & 0.00041 & Sec & DR Lyn & 54504.34187 & 0.00009 & Pri & FS Lyn & 54152.42491 & 0.00041 & Sec  \\
AA Lyn & 54091.47474 & 0.00317 & Pri & AH Lyn & 54437.63363 & 0.00047 & Sec & DR Lyn & 54525.74135 & 0.00142 & Pri & FS Lyn & 54153.50144 & 0.00039 & Sec  \\
AA Lyn & 54092.59556 & 0.00061 & Pri & AH Lyn & 54438.64814 & 0.00032 & Sec & DR Lyn & 54526.61569 & 0.00156 & Sec & FS Lyn & 54154.31478 & 0.00050 & Pri  \\
AA Lyn & 54098.77504 & 0.00032 & Pri & AH Lyn & 54439.66328 & 0.00065 & Sec & DR Lyn & 54527.50664 & 0.00038 & Pri & FS Lyn & 54154.58056 & 0.00026 & Sec  \\
AA Lyn & 54101.57519 & 0.00043 & Pri & AH Lyn & 54491.50509 & 0.00073 & Sec & DR Lyn & 54534.63023 & 0.00073 & Pri & FS Lyn & 54155.39500 & 0.00018 & Pri  \\
AA Lyn & 54111.68277 & 0.00041 & Pri & AH Lyn & 54500.65333 & 0.00087 & Sec & DR Lyn & 54535.51456 & 0.00194 & Sec & FS Lyn & 54155.66530 & 0.00031 & Sec  \\
AA Lyn & 54114.48227 & 0.00048 & Pri & AH Lyn & 54501.65906 & 0.00054 & Sec & DR Lyn & 54536.41409 & 0.00005 & Pri & FS Lyn & 54156.47532 & 0.00029 & Pri  \\
AA Lyn & 54115.61175 & 0.00073 & Pri & AH Lyn & 54502.67922 & 0.00068 & Sec & DR Lyn & 54553.34978 & 0.00200 & Sec & FS Lyn & 54157.28157 & 0.00005 & Sec  \\
AA Lyn & 54120.66564 & 0.00043 & Pri & AH Lyn & 54524.53568 & 0.00029 & Pri & DR Lyn & 54554.21918 & 0.00129 & Pri & FS Lyn & 54157.55467 & 0.00006 & Pri  \\
AA Lyn & 54123.47307 & 0.00046 & Pri & AH Lyn & 54526.56859 & 0.00102 & Pri & DR Lyn & 54558.65951 & 0.00206 & Sec & FS Lyn & 54158.38473 & 0.00033 & Sec  \\
AA Lyn & 54141.43916 & 0.00062 & Pri & AH Lyn & 54527.58434 & 0.00053 & Pri & EK Lyn & 54068.68182 & 0.00046 & Pri & FS Lyn & 54163.49561 & 0.00027 & Pri  \\
AA Lyn & 54142.55899 & 0.00096 & Pri & AH Lyn & 54530.62859 & 0.00035 & Pri & EK Lyn & 54115.63139 & 0.00117 & Pri & FS Lyn & 54165.38702 & 0.00039 & Sec  \\
AA Lyn & 54145.36689 & 0.00044 & Pri & AH Lyn & 54532.66436 & 0.00095 & Pri & EK Lyn & 54142.45463 & 0.00112 & Pri & FS Lyn & 54165.65760 & 0.00022 & Pri  \\
AA Lyn & 54146.49071 & 0.00089 & Pri & AH Lyn & 54539.27137 & 0.00175 & Sec & EK Lyn & 54151.40394 & 0.00200 & Pri & FS Lyn & 54166.46405 & 0.00043 & Sec  \\
AA Lyn & 54147.61472 & 0.00101 & Pri & AH Lyn & 54544.35038 & 0.00026 & Sec & EK Lyn & 54153.63369 & 0.00042 & Pri & FS Lyn & 54167.28921 & 0.00022 & Pri  \\
AA Lyn & 54150.42093 & 0.00019 & Pri & AH Lyn & 54547.40413 & 0.00051 & Sec & EK Lyn & 54162.58002 & 0.00096 & Pri & FS Lyn & 54167.54646 & 0.00041 & Sec  \\
AA Lyn & 54154.34954 & 0.00026 & Pri & AH Lyn & 54553.50085 & 0.00055 & Sec & EK Lyn & 54169.27849 & 0.00278 & Pri & FS Lyn & 54168.35674 & 0.00022 & Pri  \\
AA Lyn & 54155.47350 & 0.00039 & Pri & AH Lyn & 54555.53473 & 0.00071 & Sec & EK Lyn & 54171.51703 & 0.00041 & Pri & FS Lyn & 54169.43560 & 0.00034 & Pri  \\
AA Lyn & 54156.60471 & 0.00042 & Pri & AH Lyn & 54556.54901 & 0.00043 & Sec & EK Lyn & 54419.65832 & 0.00175 & Pri & FS Lyn & 54170.25455 & 0.00048 & Sec  \\
AA Lyn & 54163.33927 & 0.00062 & Pri & AH Lyn & 54557.56006 & 0.00393 & Sec & EK Lyn & 54437.54808 & 0.00451 & Pri & FS Lyn & 54170.51449 & 0.00039 & Pri  \\
AA Lyn & 54164.45003 & 0.00046 & Pri & AH Lyn & 54558.58358 & 0.00149 & Sec & EK Lyn & 54502.37924 & 0.00020 & Pri & FS Lyn & 54171.33025 & 0.00022 & Sec  \\
AA Lyn & 54168.38779 & 0.00054 & Pri & CD Lyn & 54406.71262 & 0.00123 & Pri & EK Lyn & 54504.61757 & 0.00196 & Pri & FS Lyn & 54171.59637 & 0.00048 & Pri  \\
AA Lyn & 54169.50637 & 0.00034 & Pri & CD Lyn & 54438.55591 & 0.00128 & Pri & EK Lyn & 54540.39085 & 0.00220 & Pri & FS Lyn & 54172.40493 & 0.00037 & Sec  \\
AA Lyn & 54418.75738 & 0.00066 & Pri & CD Lyn & 54447.65708 & 0.00219 & Pri & FS Lyn & 54066.57305 & 0.00031 & Sec & FS Lyn & 54192.38935 & 0.00035 & Sec  \\
AA Lyn & 54436.72370 & 0.00061 & Pri & CD Lyn & 54504.52460 & 0.00028 & Pri & FS Lyn & 54066.83345 & 0.00014 & Pri & FS Lyn & 54194.27141 & 0.00053 & Pri  \\
AA Lyn & 54439.53016 & 0.00052 & Pri & CD Lyn & 54527.26939 & 0.00108 & Pri & FS Lyn & 54067.63690 & 0.00051 & Sec & FS Lyn & 54194.55869 & 0.00047 & Sec  \\
AA Lyn & 54497.35876 & 0.00038 & Pri & CD Lyn & 54536.36740 & 0.00074 & Pri & FS Lyn & 54068.72856 & 0.00029 & Sec & FS Lyn & 54195.35491 & 0.00005 & Pri  \\
AA Lyn & 54502.40768 & 0.00043 & Pri & CD Lyn & 54545.46518 & 0.00068 & Pri & FS Lyn & 54069.55337 & 0.00068 & Pri & FS Lyn & 54436.74227 & 0.00049 & Pri  \\
AA Lyn & 54503.52984 & 0.00041 & Pri & CD Lyn & 54554.56601 & 0.00434 & Pri & FS Lyn & 54069.79863 & 0.00031 & Sec & FS Lyn & 54438.62536 & 0.00015 & Sec  \\
AA Lyn & 54526.55566 & 0.00064 & Pri & CF Lyn & 54056.53756 & 0.00038 & Sec & FS Lyn & 54070.61459 & 0.00080 & Pri & FS Lyn & 54502.61854 & 0.00009 & Pri  \\
AA Lyn & 54534.40878 & 0.00018 & Pri & CF Lyn & 54067.62583 & 0.00149 & Sec & FS Lyn & 54075.47837 & 0.00061 & Pri & FS Lyn & 54503.42752 & 0.00064 & Sec  \\
AA Lyn & 54539.45969 & 0.00065 & Pri & CF Lyn & 54069.69686 & 0.00041 & Pri & FS Lyn & 54075.74400 & 0.00036 & Sec & FS Lyn & 54523.40408 & 0.00022 & Sec  \\
AH Lyn & 54066.64114 & 0.00078 & Sec & CF Lyn & 54083.55231 & 0.00073 & Pri & FS Lyn & 54083.57324 & 0.00041 & Pri & FS Lyn & 54524.48726 & 0.00010 & Sec  \\
AH Lyn & 54067.66001 & 0.00005 & Sec & CF Lyn & 54085.63016 & 0.00069 & Sec & FS Lyn & 54083.85424 & 0.00028 & Sec & FS Lyn & 54525.30379 & 0.00012 & Pri  \\
AH Lyn & 54068.67771 & 0.00031 & Sec & CF Lyn & 54092.56599 & 0.00092 & Sec & FS Lyn & 54084.65282 & 0.00016 & Pri & FS Lyn & 54525.55949 & 0.00026 & Sec  \\
AH Lyn & 54069.69312 & 0.00014 & Sec & CF Lyn & 54094.63253 & 0.00050 & Pri & FS Lyn & 54085.73491 & 0.00020 & Pri & FS Lyn & 54526.38279 & 0.00013 & Pri  \\
AH Lyn & 54070.70987 & 0.00036 & Sec & CF Lyn & 54098.78925 & 0.00076 & Pri & FS Lyn & 54091.67340 & 0.00032 & Pri & FS Lyn & 54527.45214 & 0.00053 & Pri  \\
AH Lyn & 54074.77496 & 0.00071 & Sec & CF Lyn & 54099.49312 & 0.00115 & Sec & FS Lyn & 54092.75563 & 0.00062 & Pri & FS Lyn & 54531.49885 & 0.00038 & Sec  \\
AH Lyn & 54075.78529 & 0.00091 & Sec & CF Lyn & 54101.55962 & 0.00048 & Pri & FS Lyn & 54094.63657 & 0.00037 & Sec & FS Lyn & 54532.31317 & 0.00015 & Pri  \\
AH Lyn & 54091.54683 & 0.00007 & Pri & CF Lyn & 54114.71451 & 0.00240 & Sec & FS Lyn & 54098.69201 & 0.00027 & Pri & FS Lyn & 54532.58127 & 0.00005 & Sec  \\
AH Lyn & 54092.56149 & 0.00110 & Pri & CF Lyn & 54121.64584 & 0.00143 & Sec & FS Lyn & 54099.51019 & 0.00044 & Sec & FS Lyn & 54535.55866 & 0.00041 & Pri  \\
AH Lyn & 54094.59746 & 0.00077 & Pri & CF Lyn & 54123.72430 & 0.00208 & Pri & FS Lyn & 54099.77514 & 0.00031 & Pri & FS Lyn & 54536.36826 & 0.00048 & Sec  \\
AH Lyn & 54098.66150 & 0.00025 & Pri & CF Lyn & 54135.49908 & 0.00011 & Sec & FS Lyn & 54100.57688 & 0.00064 & Sec & FS Lyn & 54536.63994 & 0.00035 & Pri  \\
AH Lyn & 54099.67648 & 0.00061 & Pri & CF Lyn & 54139.65437 & 0.00059 & Sec & FS Lyn & 54100.84732 & 0.00021 & Pri & FS Lyn & 54537.44138 & 0.00020 & Sec  \\
AH Lyn & 54100.69420 & 0.00030 & Pri & CF Lyn & 54140.35021 & 0.00040 & Pri & FS Lyn & 54101.66338 & 0.00039 & Sec & FS Lyn & 54539.33791 & 0.00020 & Pri  \\
AH Lyn & 54101.71020 & 0.00023 & Pri & CF Lyn & 54142.42875 & 0.00110 & Sec & FS Lyn & 54111.65288 & 0.00054 & Pri & FS Lyn & 54539.60308 & 0.00019 & Sec  \\
AH Lyn & 54115.43563 & 0.00146 & Sec & CF Lyn & 54146.58725 & 0.00099 & Sec & FS Lyn & 54114.62174 & 0.00039 & Sec & FS Lyn & 54540.41809 & 0.00058 & Pri  \\
AH Lyn & 54116.45495 & 0.00091 & Sec & CF Lyn & 54153.51475 & 0.00078 & Sec & FS Lyn & 54115.43699 & 0.00060 & Pri & FS Lyn & 54541.49754 & 0.00015 & Pri  \\
AH Lyn & 54118.47987 & 0.00080 & Sec & CF Lyn & 54155.59123 & 0.00005 & Pri & FS Lyn & 54115.69815 & 0.00043 & Sec & FS Lyn & 54544.46391 & 0.00033 & Sec  \\
AH Lyn & 54120.51090 & 0.00077 & Sec & CF Lyn & 54162.51781 & 0.00027 & Pri & FS Lyn & 54116.51968 & 0.00032 & Pri & FS Lyn & 54545.55342 & 0.00022 & Sec  \\
AH Lyn & 54121.53030 & 0.00063 & Sec & CF Lyn & 54165.29099 & 0.00056 & Pri & FS Lyn & 54118.40445 & 0.00039 & Sec & FS Lyn & 54553.38163 & 0.00051 & Pri  \\
AH Lyn & 54122.54438 & 0.00104 & Sec & CF Lyn & 54167.36922 & 0.00088 & Sec & FS Lyn & 54118.67406 & 0.00020 & Pri & FS Lyn & 54554.45754 & 0.00034 & Pri  \\
AH Lyn & 54123.55731 & 0.00088 & Sec & CF Lyn & 54169.44452 & 0.00084 & Pri & FS Lyn & 54120.56704 & 0.00031 & Sec & FS Lyn & 54555.54059 & 0.00028 & Pri  \\
AH Lyn & 54141.35516 & 0.00011 & Pri & CF Lyn & 54171.51833 & 0.00071 & Sec & FS Lyn & 54121.64454 & 0.00037 & Sec & FS Lyn & 54556.36089 & 0.00005 & Sec  \\
AH Lyn & 54142.37156 & 0.00069 & Pri & CF Lyn & 54436.81879 & 0.00028 & Pri & FS Lyn & 54122.45438 & 0.00045 & Pri & FS Lyn & 54556.60870 & 0.00082 & Pri  \\
AH Lyn & 54143.38332 & 0.00053 & Pri & CF Lyn & 54439.58947 & 0.00091 & Pri & FS Lyn & 54122.72509 & 0.00054 & Sec & FS Lyn & 54557.43236 & 0.00048 & Sec  \\
AH Lyn & 54145.41546 & 0.00061 & Pri & CF Lyn & 54500.53616 & 0.00077 & Pri & FS Lyn & 54123.53178 & 0.00011 & Pri & FS Lyn & 54558.50803 & 0.00046 & Sec  \\
AH Lyn & 54146.43718 & 0.00048 & Pri & CF Lyn & 54523.39820 & 0.00052 & Sec & FS Lyn & 54139.46764 & 0.00043 & Sec & FS Lyn & 54573.34878 & 0.00028 & Pri  \\
AH Lyn & 54147.44563 & 0.00021 & Pri & CF Lyn & 54525.47777 & 0.00105 & Pri & FS Lyn & 54139.74027 & 0.00051 & Pri & FS Lyn & 54574.44005 & 0.00023 & Pri  \\
 \hline
\end{tabular}}
\end{table}

\end{document}

%% file: table1.tex
 \tiny
\begin{tabular}{ c c c c c c c}
\hline \hline
 Parameter     &    RV Lyn                &    AA Lyn               &      AH Lyn               &    CD Lyn                 \\ \hline
 $JD_0-2400000$& 54409.5592 $\pm$ 0.0005  & 54056.6671 $\pm$ 0.0008 & 54091.5447 $\pm$ 0.0010   & 54504.5198 $\pm$ 0.0004   \\
 $P$ [d]       & 2.307640 $\pm$ 0.0000004 & 0.5613884 $\pm$ 0.0000003& 1.01641142 $\pm$ 0.000008& 2.2747194 $\pm$ 0.000008  \\
 $i$ [deg]     & 84.91 $\pm$ 0.40         & 72.82 $\pm$ 0.46        &  88.75 $\pm$ 0.42         &  81.93 $\pm$ 0.70         \\
 $q = M_2/M_1$ & 0.69 $\pm$  0.02         & 0.87 $\pm$ 0.04         &  0.86 $\pm$ 0.03          &  0.53 $\pm$ 0.04          \\
 Type          &   Detached               & Semidetached            &  Detached                 &  Detached                 \\
 $T_1$ [K]     & 6700 (fixed)             & 5600 (fixed)            &  6500 (fixed)             &  6800 (fixed)             \\
 $T_2$ [K]     & 4287 $\pm$ 80            & 3760 $\pm$ 120          &  6074 $\pm$ 52            &  4350 $\pm$ 48            \\
 $\Omega_1$    & 5.243 $\pm$ 0.081        & 4.984 $\pm$ 0.118       & 5.455 $\pm$ 0.037         &  4.946 $\pm$ 0.016        \\
 $\Omega_2$    & 3.785 $\pm$ 0.024        & 3.535 $^b$              & 5.588 $\pm$ 0.035         &  4.502 $\pm$ 0.013        \\
 $L_1/(L_1+L_2)$ [\%] & 88.8 $\pm$ 1.3    & 86.3 $\pm$ 4.5          & 64.3 $\pm$ 1.3            &  94.4 $\pm$ 2.7           \\
 $L_2/(L_1+L_2)$ [\%] & 11.2 $\pm$ 0.8    & 13.7 $\pm$ 1.3          & 35.7 $\pm$ 1.1            &   5.6 $\pm$ 2.1           \\
 $l_3$ [\%] $^a$    &  0.0                &  7.6 $\pm$ 2.1          &  0.0                      &   0.0                     \\
 $R_1/a$       &  0.210 $\pm$ 0.011       & 0.253 $\pm$ 0.021       &  0.219 $\pm$ 0.008        &  0.228 $\pm$ 0.040        \\
 $R_2/a$       &  0.268 $\pm$ 0.010       & 0.376 $\pm$ 0.018       &  0.191 $\pm$ 0.007        &  0.163 $\pm$ 0.031        \\ \hline
 Spot Lat. [deg]  &  --                   &   --                    &   --                      &  84.7 $\pm$ 1.8  \\
 Spot Long. [deg] &  --                   &   --                    &   --                      & 328.8 $\pm$ 0.4  \\
 Spot Rad. [deg]  &  --                   &   --                    &   --                      &  13.6 $\pm$ 1.2  \\
 Spot Temp.       &  --                   &   --                    &   --                      &  0.82 $\pm$ 0.04 \\ \hline
\end{tabular}

%% file: table2.tex
 \tiny
\begin{tabular}{ c c c c c c c}
\hline \hline
 Parameter     &    CF Lyn                &     DR Lyn              &      EK Lyn              &     FS Lyn                \\ \hline
 $JD_0-2400000$& 54069.6974 $\pm$ 0.0007  & 54502.5720 $\pm$ 0.0026 & 54068.6844 $\pm$ 0.0005  & 54142.4354 $\pm$ 0.0003   \\
 $P$ [d]       & 1.3853727 $\pm$ 0.0000029&1.7808806 $\pm$ 0.0000011& 2.2355353 $\pm$ 0.000093 & 0.5400052 $\pm$ 0.0000030 \\
 $i$ [deg]     & 86.31 $\pm$ 0.79         & 85.96 $\pm$ 0.69        &  83.13 $\pm$ 0.98        &  63.50 $\pm$ 0.94         \\
 $q = M_2/M_1$ & 0.85 $\pm$ 0.06          & 0.74 $\pm$ 0.02         &  0.42 $\pm$ 0.03         &  0.71 $\pm$ 0.09          \\
 Type          & Detached                 & Detached                &  Detached                &  Detached                 \\
 $T_1$ [K]     & 6150 (fixed)             & 6690 (fixed)            &  8840 (fixed)            &  7100 (fixed)             \\
 $T_2$ [K]     & 5100 $\pm$ 72            & 4370 $\pm$ 72           &  5325 $\pm$ 110          &  5019 $\pm$ 66            \\
 $\Omega_1$    & 4.209 $\pm$ 0.022        & 5.100 $\pm$ 0.027       & 4.828 $\pm$ 0.040        &  3.292 $\pm$ 0.025        \\
 $\Omega_2$    & 7.269 $\pm$ 0.046        & 3.580 $\pm$ 0.010       & 4.533 $\pm$ 0.027        &  3.297 $\pm$ 0.029        \\
 $L_1/(L_1+L_2)$ [\%] & 92.5 $\pm$ 0.8    & 84.6 $\pm$ 0.8          & 96.9 $\pm$ 3.7           &  87.2 $\pm$ 1.7           \\
 $L_2/(L_1+L_2)$ [\%] &  7.5 $\pm$ 0.3    & 15.4 $\pm$ 0.4          &  3.1 $\pm$ 1.2           &  12.8 $\pm$ 0.9           \\
 $l_3$ [\%] $^a$    & 18.1 $\pm$ 1.0      &  0.0                    &  3.7 $\pm$ 0.7           &   0.0                     \\
 $R_1/a$       &  0.305 $\pm$ 0.012       & 0.231 $\pm$ 0.011       &  0.228 $\pm$ 0.021       &  0.402 $\pm$ 0.027        \\
 $R_2/a$       &  0.140 $\pm$ 0.009       & 0.310 $\pm$ 0.008       &  0.127 $\pm$ 0.034       &  0.342 $\pm$ 0.024        \\ \hline
\end{tabular}